\documentclass[aps,prb,reprint,citeautoscript,superscriptaddress,showpacs]{revtex4-1}

\usepackage{graphicx,epsfig,float}
\usepackage{amssymb}
\usepackage{amsmath}

\begin{document}
\title{Reconstruction of Zigzag Graphene Edges: Energetics, Kinetics and Residual Defects}

\author{Yulia G. Polynskaya}
\email{yupol@kintechlab.com}
\affiliation{Kintech Lab Ltd., 3rd Khoroshevskaya Street 12, Moscow 123298, Russia}
\author{Irina V. Lebedeva}
\email{liv\_ira@hotmail.com}
\affiliation{CIC nanoGUNE BRTA, San Sebasti\'an 20018, Spain}
\affiliation{Catalan Institute of Nanoscience and Nanotechnology - ICN2, CSIC and BIST, Campus UAB, Bellaterra 08193, Spain}
\affiliation{Simune Atomistics, Avenida de Tolosa 76, San Sebasti\'an 20018, Spain}
\author{Andrey A. Knizhnik}
\email{knizhnik@kintechlab.com}
\affiliation{Kintech Lab Ltd., 3rd Khoroshevskaya Street 12, Moscow 123298, Russia}
\affiliation{National Research Centre ``Kurchatov Institute", Kurchatov Square 1, Moscow 123182, Russia}
\author{Andrey M. Popov}
\email{popov-isan@mail.ru}
\affiliation{Institute for Spectroscopy of Russian Academy of Sciences, Troitsk, Moscow 108840, Russia}

\begin{abstract}
{\it Ab initio} calculations are performed to study consecutive reconstruction of a zigzag graphene edge. According to the obtained energy profile along the reaction pathway, the first reconstruction step, formation of the first pentagon-heptagon pair, is the slowest one, while the growth of an already nucleated reconstructed edge domain should occur steadily at a much higher rate. Domains merge into one only in 1/4 of cases when they get in contact, while in the rest of the cases, residual defects are left. Structure, energy and magnetic properties of these defects are studied. It is found that spontaneous formation of pairs of residual defects (i.e. spontaneous domain nucleation) in the fully reconstructed edge is unlikely at temperatures below 1000 K. Using a kinetic model, we show that the average domain length is of several $\mathrm{\mu}$m at room temperature and it decreases exponentially upon increasing the temperature at which the reconstruction takes place. 
\end{abstract}

\maketitle

The success of nanoelectronic devices based on graphene nanostructures \cite{Geim2009}, such as graphene nanoribbons (GNRs) \cite{Han2007, Ritter2009}, is conditioned by the capacity to control precisely the atomistic structure \cite{Banhart2011,Terrones2012,Skowron2015}. 
The edges are much more reactive and prone to defect formation compared to the graphene bulk \cite{Skowron2015}. At the same time, they have a drastic effect on the electronic \cite{Ritter2009, Nakada1996, Kunstmann2011, Son2006, Lebedeva2012, Lee2009, Bhandary2010, Gunlycke2007, Li2010, Koskinen2008, Tao2011, Niimi2006, Kobayashi2005}, magnetic \cite{Cheng2012, Gan2010, Son2006, Kunstmann2011, Seitsonen2010, Magda2014, Polynskaya2022}, mechanical \cite{Cheng2012prb, Gan2010, Huang2009} and chemical \cite{Gan2010, Koskinen2008, Seitsonen2010, Wassmann2008} properties of graphene nanostructures. Therefore, significant efforts have been made to study structure \cite{Gan2010, Huang2009, Kobayashi2005, Niimi2006, Koskinen2008, Gunlycke2007, Bhandary2010, Kunstmann2011, Magda2014, Wassmann2008, Girit2009, Chuvilin2009, Koskinen2009, Kim2013, Warner2014, He2015, Song2010, Lee2010, Ivanovskaya2011}, transformations \cite{Cheng2012, Cheng2012prb, Koskinen2008, Li2010, Chuvilin2009, Kim2013, He2015, Dang2017, Kroes2011, Kotakoski2012, Polynskaya2022} and defects \cite{Skowron2015, Li2010, Huang2008, Kunstmann2011, Polynskaya2022} at graphene edges.

\begin{figure}
   \centering
 \includegraphics[width=\columnwidth]{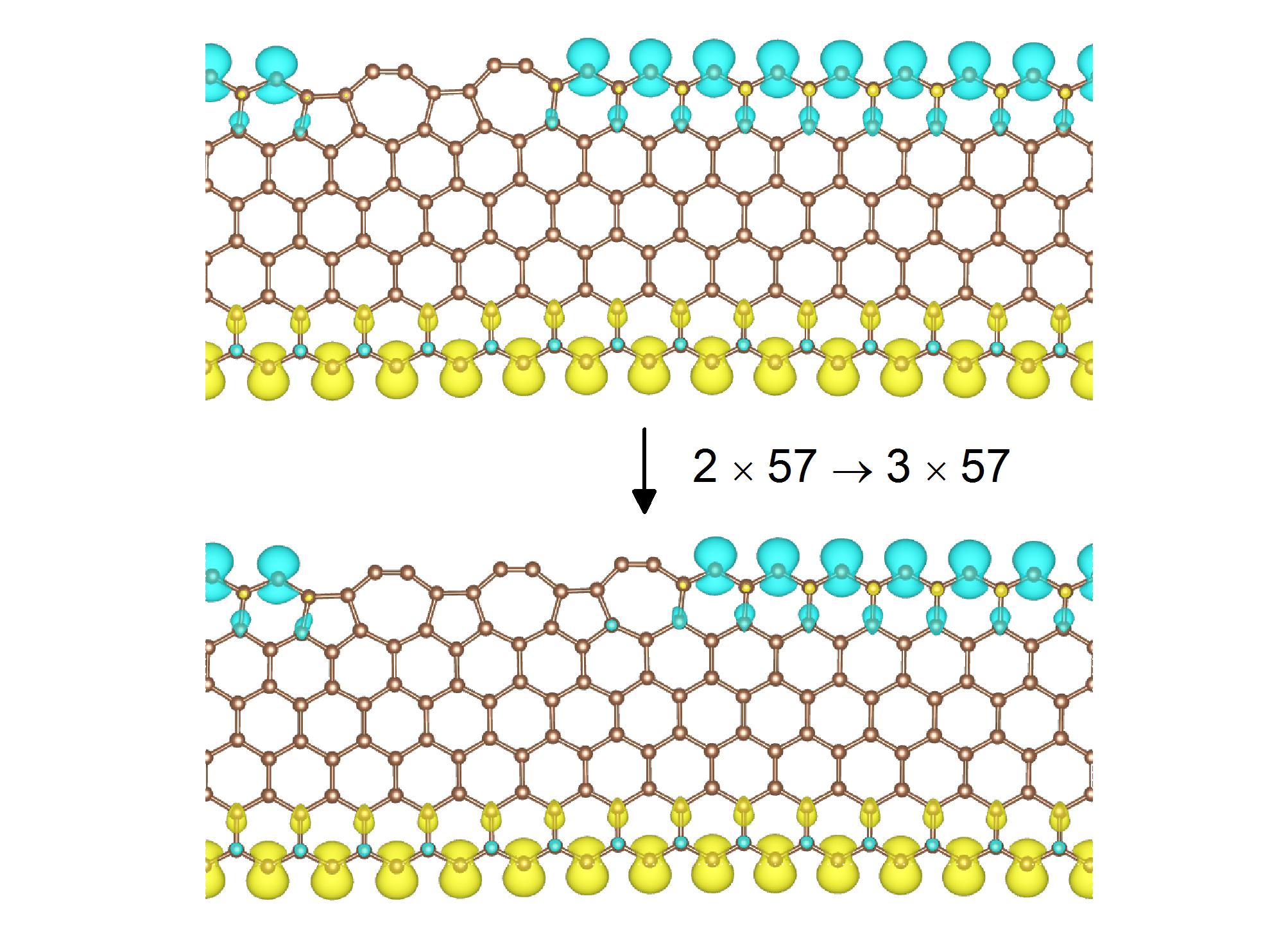}
   \caption{One simulation cell of the 6-ZGNR after the second and third steps of the zigzag edge reconstruction. Spin maps are shown (isosurfaces 0.01 $e/$\AA$^3$).} 
   \label{fig:example}
\end{figure}

One of the most famous examples of graphene edge transformations is zigzag edge reconstruction \cite{Gan2010, Huang2009, Koskinen2008, Kunstmann2011, Wassmann2008, Song2010, Lee2010,  Ivanovskaya2011, Kroes2011, Kotakoski2012}. In this process, pairs of hexagons at pristine zigzag edges are transformed into pentagon-heptagon (57) pairs (Figure \ref{fig:example}). The reconstruction leads to formation of triple bonds similar the ones at the armchair edge and thus reduces the number of dangling bonds and the energy of zigzag edges \cite{Gan2010, Huang2009, Koskinen2008, Kunstmann2011, Wassmann2008, Song2010, Lee2010,  Ivanovskaya2011, Kroes2011, Kotakoski2012}. Formation of 57 pairs suppresses magnetization of zigzag edges \cite{Cheng2012, Cheng2012prb, Kunstmann2011, Polynskaya2022} and results in a decrease of the conductance \cite{Li2010}.

In spite of being more stable thermodynamically than other pristine graphene edges, reconstructed zigzag edges are not the most abundant ones in the experiments \cite{Girit2009, Chuvilin2009, Koskinen2009, Cheng2012prb, Kim2013, Warner2014}. The reasons can be large barriers for the reconstruction \cite{Cheng2012prb, Kroes2011}, edge contamination \cite{He2015} and the effect of electron irradiation on edge stability \cite{Girit2009, Koskinen2009}. In the present Letter we consider pristine edges, which can be obtained by annealing at 600 $^{\circ}$C \cite{Cheng2012prb, He2015} or irradiation by electrons with the kinetic energy insufficient to cause carbon bond rearrangements \cite{Sinitsa2021}. This is the case for typical transmission electron microscopy studies of the reconstruction \cite{Girit2009, Chuvilin2009, Koskinen2009, Cheng2012prb, Kim2013, Warner2014}. Previous calculations \cite{Cheng2012, Cheng2012prb, Li2010, Dang2017, Polynskaya2022} for pristine graphene edges gave a significant barrier for the first reconstruction step, formation of the first 57 pair. However, the overall kinetics of the edge reconstruction depends also on consecutive generation of 57 pairs. In particular, the activation barrier, $E_\mathrm{a}$ , i.e. the energy difference between the transition and initial states, and reaction energy, $\Delta E$, i.e. the energy difference between the final and initial states, (Figure \ref{fig:example}) estimated from the experimental observations of propagation of a reconstructed edge domain ($E_\mathrm{a}\sim1.3$ eV and $\Delta E\sim-1.2$ eV from Ref. \onlinecite{Kim2013}) differ strongly from the calculation results for the first reconstruction step (e.g., $E_\mathrm{a}\sim1.6$ eV and $\Delta E\sim-0.2$ eV from Ref. \onlinecite{Polynskaya2022}). Although there were already attempts to consider energetics of several reconstruction steps from the first principles\cite{Cheng2012, Lee2010, Dang2017}, the models employed were clearly too small to obtain accurate results and to judge about how the whole process occurs. Last stages of reconstruction at which reconstructed edge domains get in contact, structure and density of residual defects have not been considered at all.

Here we carry out {\it ab initio} calculations of consecutive generation of 57 pairs at the zigzag graphene edge using an atomistic model that is sufficiently large to describe adequately all stages of the reconstruction process. Using a simple kinetic model with the parameters extracted from the {\it ab initio} calculations performed, we obtain the distribution of domains of the reconstructed zigzag edge in length and its dependence on the temperature at which the reconstruction takes place. We also consider residual defects that can be left between reconstructed edge domains and study their energetics and magnetic properties.

In our recent study \cite{Polynskaya2022}, we showed that accurate {\it ab initio} calculations of the reaction energy and activation barrier (Figure \ref{fig:example}) for  formation of the first 57 pair at the graphene edge require the following minimal nanoribbon model: the nanoribbon should consist of at least 6 zigzag rows and the distance between periodic images of pentagon-heptagon pairs along the nanoribbon axis should be at least 6 hexagons. Here we perform the calculations for the 6-ZGNR (zigzag graphene nanoribbon consisting of 6 zigzag rows) in the simulation cell including 12 hexagons along the ZGNR axis (Figure \ref{fig:example}). This allows us to properly model nucleation, growth and merging of reconstructed edge domains.

Spin-polarized density functional theory calculations have been performed using the VASP code \cite{Kresse1996} with the Perdew-Burke-Ernzerhof functional \cite{Perdew1996}. The projector-augmented wave method (PAW) \cite{Kresse1999} is applied to describe the interaction of valence and core electrons. The cutoff kinetic energy of the plane-wave basis set is 500 eV. The tolerance achieved in self-consistent iterations is $10^{-8}$ and the Gaussian smearing of width 0.05 eV is applied. Since the aim of the study is to model edge reconstruction for graphene, the elementary unit cell of the pristine ZGNR is taken equal to the lattice constant of graphene, $a_0 = 2.466$ \AA{} according to our calculations. The vacuum gap of 10 \AA{} across the ZGNR and perpendicular to ZGNR plane is introduced to minimize the interaction between periodic images of the ZGNR.  To stimulate convergence to the antiferromagnetic state, initial spins with anti-parallel ordering at the opposite ZGNR edges, respectively, are set at the edge atoms.  The Brillouin-zone integration is performed using the $3\times 1 \times1$ Monkhorst-Pack grid \cite{Monkhorst1976}. The residual atomic forces in geometry optimization do not exceed 0.003 eV/\AA{}.

The pathway for consecutive generation of 57 pairs is investigated using the nudged elastic band (NEB) method \cite{Mills1995, Jonsson1998} with 6 images between the initial and final states at each step. To reduce the computational effort, the NEB calculations are performed for the maximal kinetic energy of the plane wave basis set of 400 eV, the tolerance $10^{-4}$ and smearing width of 0.2 eV. Damped molecular dynamics with the maximal residual force of 0.03 eV/\AA{} is used to optimize the geometry of images. After the optimization, the energy of the transition state is computed with the same parameters as of the initial and final state. The spin maps are extracted and plotted using the VASPKIT \cite{Wang2021} and VESTA \cite{Momma2011}, respectively.

\begin{figure}
   \centering
 \includegraphics[width=\columnwidth]{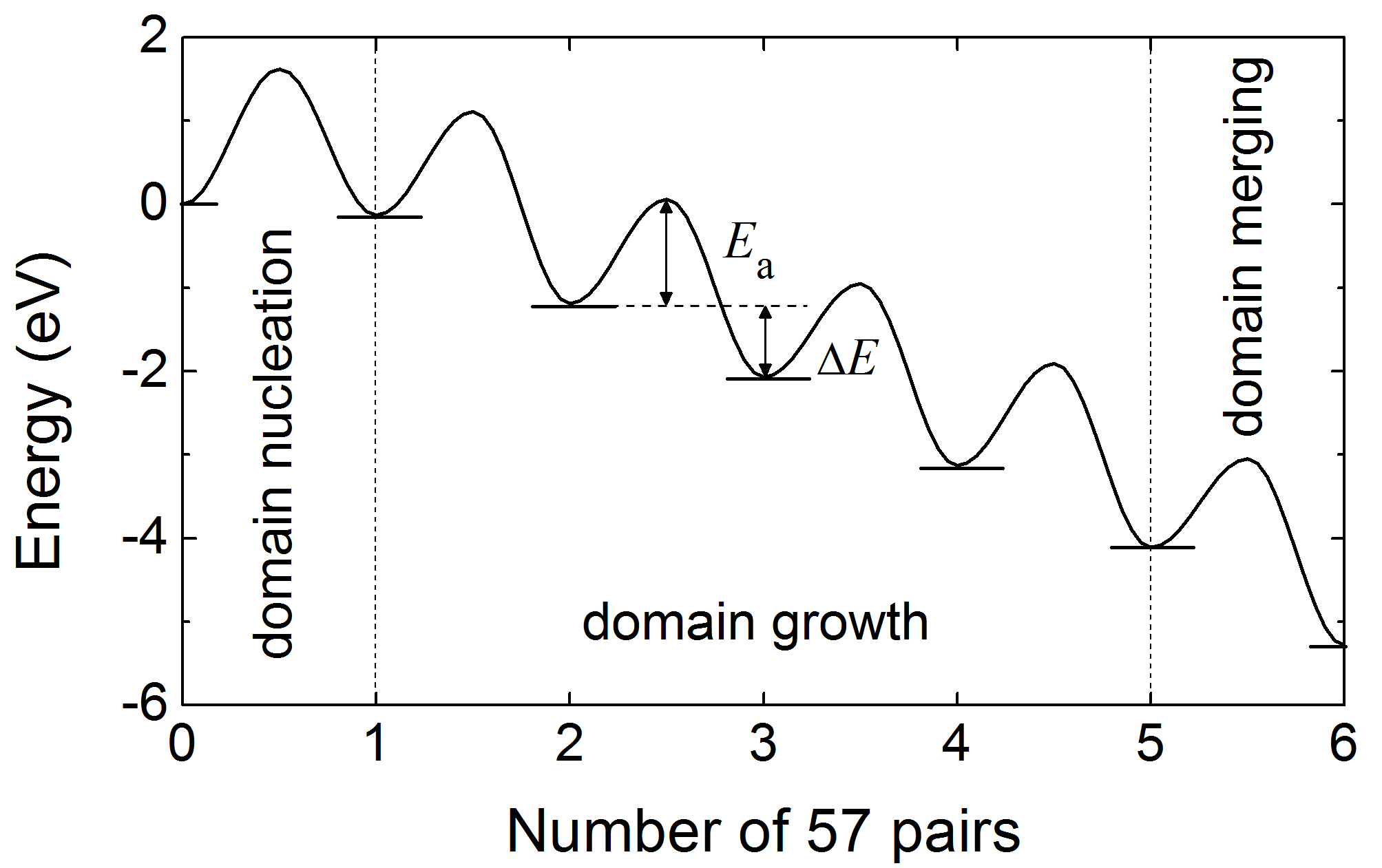}
   \caption{Energy (in eV) along the reaction pathway for consecutive reconstruction of the zigzag edge in the simulation cell including 12 hexagons along the edge. The energy is given relative to the unreconstructed edge. Stages of the evolution of reconstructed edge domains are denoted. The activation and reaction energies, $E_\mathrm{a}$ and $\Delta E$, respectively, for formation of the third 57 pair are indicated.} 
   \label{fig:energy}
\end{figure}

The computed energy profile for the zigzag edge reconstruction is shown in Figure \ref{fig:energy}. It is seen from this figure that consecutive formation of 57 pairs leads to the gradual energy release and the highest barrier for the whole reconstruction process corresponds to formation of the first 57 pair. 

The activation barriers and reaction energies for consecutive generation of 57 pairs are listed in Table \ref{table:energies}. According to these results, the activation barrier and reaction energy are reduced by $\delta E\sim0.4$ eV and $\sim0.9$ eV, respectively, when 57 pairs are formed close to the existing ones. Therefore, once the first 57 pair is formed, the following 57 pairs should predominantly arise close to it, that is a reconstructed edge domain starts to grow. 

It should be emphasized that for steps 1--3 of the reconstruction in our calculations, the periodic images of the reconstructed edge domain are separated by more than 6 hexagons and can be considered as isolated \cite{Polynskaya2022}. Steps 2 and 3 have virtually the same activation barriers ($\sim1.2$ eV) and reaction energies ($\sim-1$ eV)  indicating that they correspond to a steady domain growth. Note that these kinetic parameters are in good agreement with the estimates $E_\mathrm{a}\sim1.3$ eV and $\Delta E\sim-1.2$ eV from the experimental frequencies of formation and destruction of 57 pairs\cite{Kim2013}. At steps 4--6 in our calculations, the periodic images of the reconstructed edge domain get close. Therefore, they describe the situation when two adjacent domains approach each other. Still we find that steps 4 and 5 have the reaction energies and activation barriers close to those for steps 2 and 3 and thus they also should be attributed to the growth stage. Step 6 has a noticeably smaller activation barrier and reaction energy compared to the previous steps and corresponds to domain merging.

\begin{table}
    \caption{Activation and reaction energies, $E_\mathrm{a}$ and $\Delta E$, respectively, computed for consecutive steps of zigzag edge reconstruction in the simulation cell including 12 hexagons along the edge (stages of evolution of reconstructed edge domains are indicated).}
   \renewcommand{\arraystretch}{1.2}
   \setlength{\tabcolsep}{12pt}
    \resizebox{\columnwidth}{!}{
        \begin{tabular}{*{4}{c}}
\hline
stage & 57 pair number & $E_\mathrm{a}$  (eV)  & $\Delta E$  (eV) \\\hline
nucleation & 1 & 1.62  & -0.13  \\\hline
growth & 2 & 1.24  & -1.06  \\
&3 & 1.25  & -0.88  \\
&4 & 1.12  & -1.06  \\
&5 & 1.22  & -0.98  \\\hline
merging &6 & 1.06  & -1.17  \\\hline
\end{tabular}
}
\label{table:energies}
\end{table}

Based on the Arrhenius equation, the ratio of the kinetic constants for domain nucleation and growth is:
\begin{equation} \label{eq_constants}
  \frac{k_\mathrm{nucl}}{k_\mathrm{gr}} = \exp\left(-\frac{\delta E}{k_\mathrm{B}T}\right) ,
\end{equation}
i.e. nucleation is much slower than domain growth ($k_\mathrm{nucl}/k_\mathrm{gr}\sim2\cdot10^{-7}$ at room temperature). Analogously, formation of the very last 57 pair leading to domain merging should occur faster than the steady domain growth.

\begin{figure}
   \centering
 \includegraphics[width=\columnwidth]{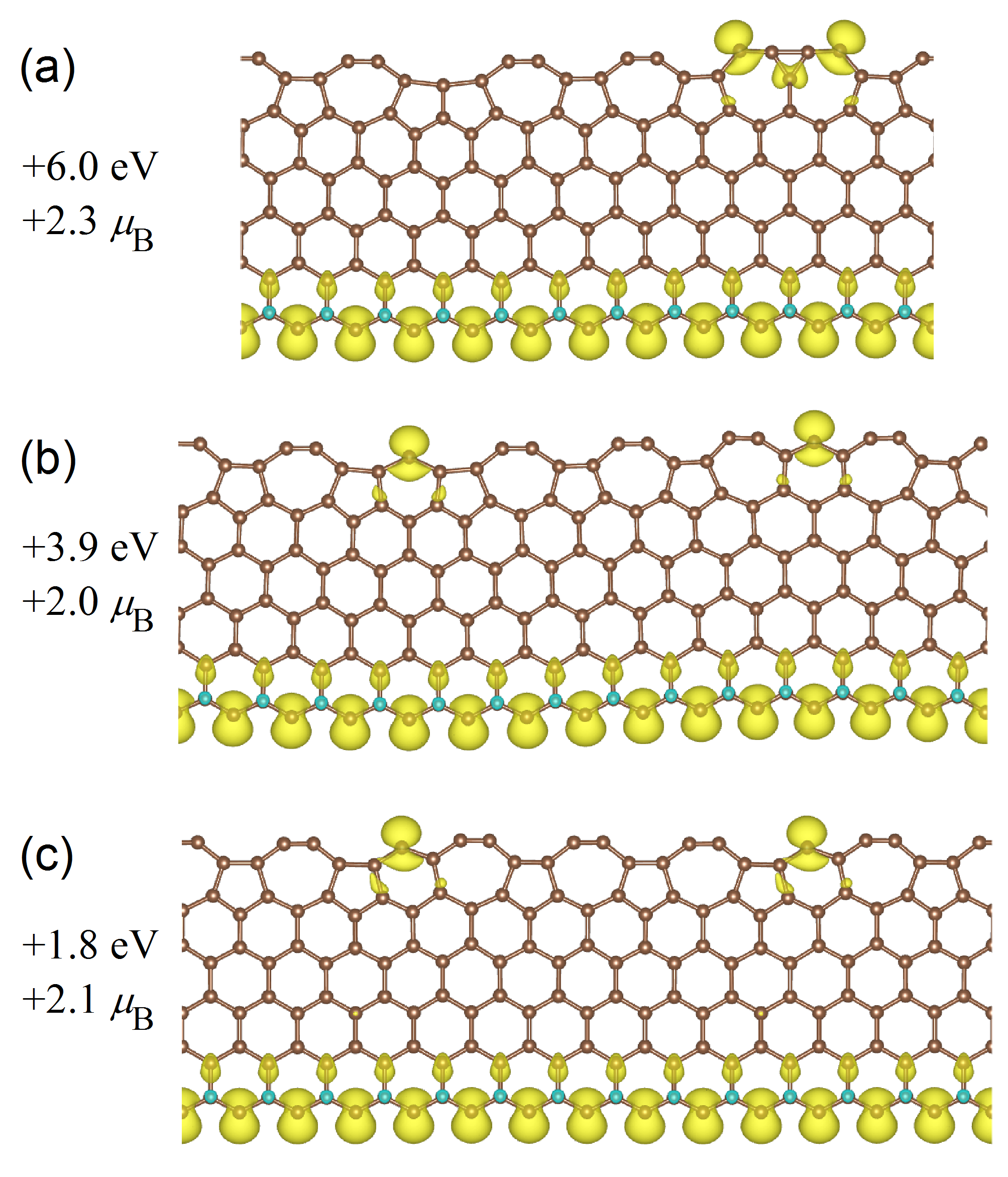}
   \caption{One simulation cell of the 6-ZGNR with a fully reconstructed edge and different boundaries (residual defects) between reconstructed edge domains: (a) 55 and 77, (b) 767 and 565, and (c) a 765 pair. Relative energies and magnetic moments of the structures are given with respect to the fully reconstructed edge without residual defects. Spin maps are shown (isosurfaces 0.01 $e/$\AA$^3$). } 
   \label{fig:boundaries}
\end{figure}

It should be noted that merging of two reconstructed edge domains into one is possible only if (1) the domains have the same orientation, i.e. 57 pairs are oriented in the same way, and (2) the number of hexagons separating domains is even. This corresponds only to 1/4 of cases of domains getting in contact. In the opposite 3/4 of cases, domains are separated by residual defects that play the role of domain boundaries. Spontaneous formation of such defects at the reconstructed edge (as well as consideration of these defects under periodic boundary conditions) is possible only in pairs like the ones shown in  Figure \ref{fig:boundaries}. During the edge reconstruction, boundaries between growing domains  are formed independently from each other.  

For 1/4 of domains getting in contact, the domains are counter-aligned and the number of hexagons between them is even. The shortest domain boundaries in this case correspond to 55 and 77 defects (Figure \ref{fig:boundaries}a). However, the calculations show that the formation energy of a pair of such residual defects, i.e. the relative energy as compared to the fully reconstructed edge without domain boundaries, is very high because of the significant deformation of carbon rings in the defects. In this case it is actually energetically favourable to leave hexagons between the domains:  the energy is reduced by almost 2 eV if the domain boundaries consist of 5665 and 7667 defects instead of 55 and 77. A similar formation energy of defect pairs of $\sim4$ eV is found for 1/4 of cases when the domains are counter-aligned and the number of hexagons between them is odd. Then the most energetically favourable pair of residual defects is 767 and 565  (Figure \ref{fig:boundaries}b). In 1/4 of cases when domains are co-aligned and the number of hexagons between them is odd, the domain boundaries correspond to 765 defects (Figure \ref{fig:boundaries}c). The formation energy of the 765 defect pair of 1.8 eV is the smallest among the possible domain boundaries. Still it is too high to be able to observe spontaneous formation of such defects, i.e. spontaneous domain nucleation, in the reconstructed edge at temperatures below 1000 K. 

According to our calculations, all the structures with domain boundaries are flat, the same as in their absence. We also find that edge atoms at the boundaries carry significant magnetic moments $\sim\mu_\mathrm{B}$ (Bohr magneton) parallel to those at the opposite nanoribbon edge (Figure \ref{fig:boundaries}). Similar magnetic moments were obtained previously for short regions of the unreconstructed zigzag edge\cite{Polynskaya2022}. 

To model domain formation at different temperatures, we use a simple kinetic model.  A periodic array of $N$ hexagons at the unreconstructed zigzag edge is considered. The probability of nucleation of a new domain of the reconstructed edge at each time step of duration $\delta t = k_\mathrm{gr}^{-1}$ is $p =k_\mathrm{nucl}/k_\mathrm{gr}$ (see Eq. (\ref{eq_constants})) for each pair of adjacent hexagons. At each time step, each reconstructed edge domain grows by two hexagons at each end if there is a place to grow. When 2 or 3 hexagons are left between adjacent domains, it is selected randomly which one of them grows. In this Letter we restrict ourselves to consideration of formation of the initial domain structure and the simulation finishes when there is no room left for domains to grow or nucleate. Migration and annihilation of residual defects are neglected. Migration events should not normally be accompanied by a significant energy change and can occur with virtually equal probabilities in both directions (contrary to motion of domain boundaries during the domain growth). Thus, migration  that can finally result in annihilation of residual defects should be slow compared to domain growth. Changes in the domain structure after the full edge reconstruction due to such processes will be studied elsewhere.

\begin{figure}
   \centering
 \includegraphics[width=\columnwidth]{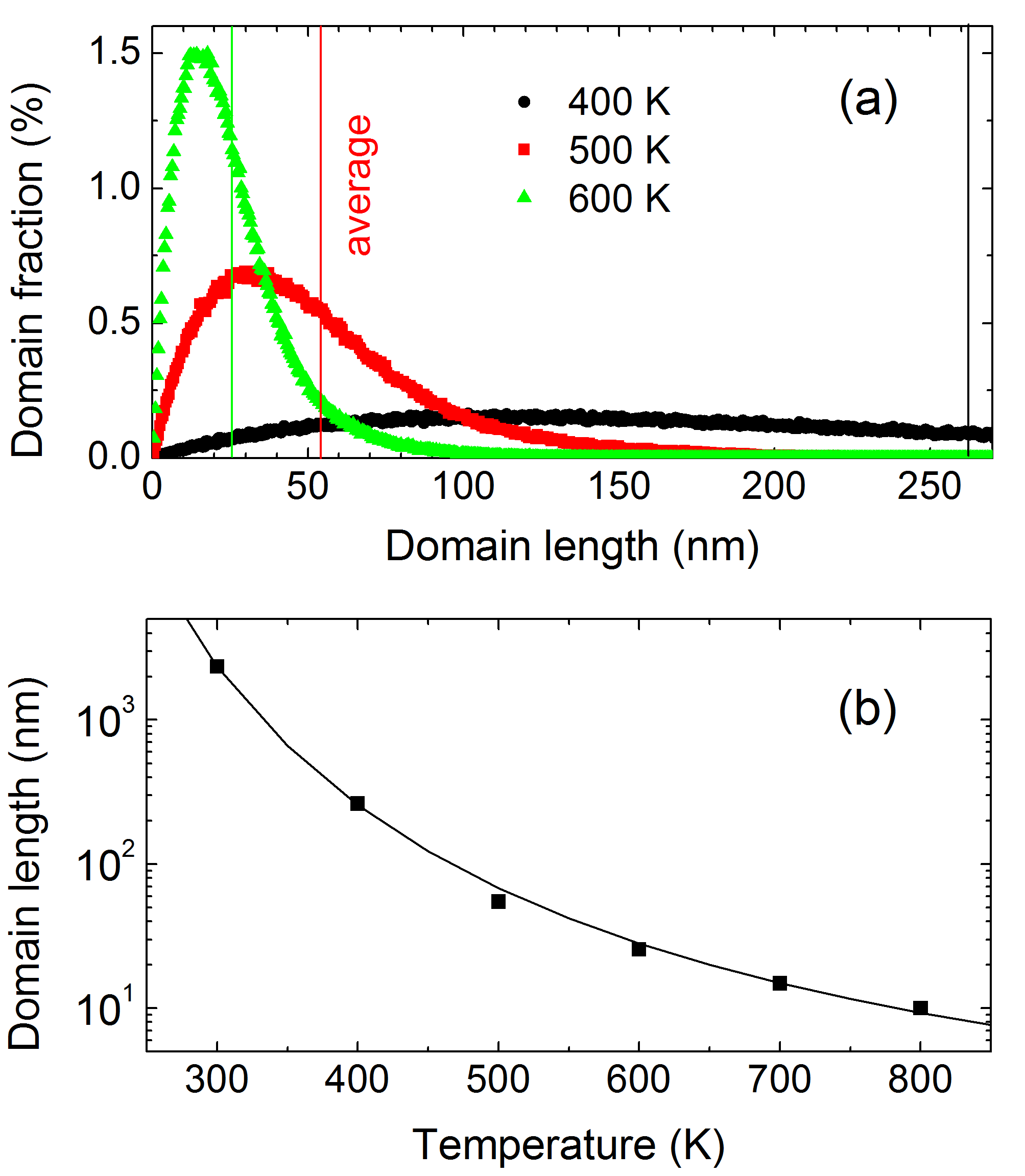}
   \caption{(a) Distributions of lengths of reconstructed edge domains (in nm) obtained using the kinetic model at different temperatures during the reconstruction: (black circles) 400 K, (red squares) 500 K and (green triangles) 600 K. The average domain lengths are indicated by the vertical lines. (b) Calculated average length of reconstructed edge domains (in nm) as a function of temperature (in K). The exponential approximation according to Eq. (\ref{approx}) is shown by the solid line.} 
   \label{fig:length}
\end{figure}

The calculations have been performed for edges of length up to $N = 3\cdot 10^6$ hexagons. The distribution of domain lengths has been obtained based on 10--1000 calculations for each temperature. Examples of such a distribution at different temperatures are shown in Figure \ref{fig:length}a. They are obtained based on the data for more than 400000 domains at each temperature. The average length is about twice greater than the position of the maximum at the same temperature. Upon increasing temperature, the maximum and average points are shifted to smaller lengths. 

The dependence of the average domain length on the temperature during the reconstruction is given in Figure \ref{fig:length}b. At room temperature, the average domain length exceeds 2 $\mathrm{\mu}$m, while at 800 K it goes down to 10 nm. The dependence can be approximated by an exponential law:
\begin{equation} \label{approx}
  L = (0.34\pm0.03)\mathrm{[nm]} \exp{\left(-\frac{0.229\pm0.003\mathrm{[eV]}}{k_\mathrm{B}T}\right)}
\end{equation}
Note that the energy in the exponent here is almost twice smaller than that in the ratio of the rates for domain nucleation and growth (see Eq. (\ref{eq_constants})). The absence of a simple relation between these two energy factors in the exponent should be a consequence of the complexity of the whole reconstruction process in which domains can be nucleated at different time moments and also can merge under the conditions discussed before. Analytic justification of this approximation is beyond the scope of the present Letter.

To summarize, our {\it ab initio} calculations show that the zigzag edge reconstruction can be divided into three stages: nucleation, steady growth and getting in contact of reconstructed edge domains. The growth of already nucleated domains occurs at a much faster rate than nucleation. The difference in the barriers at these two stages is $\sim0.4$ eV. Domains easily merge into one in 1/4 of cases when domains get in contact. In the rest 3/4 of cases, residual defects are left. These defects carry magnetic moments  $\sim\mu_\mathrm{B}$  and increase significantly the system energy as compared to the reconstructed edge without residual defects. The average length of reconstructed edge domains is of several $\mathrm{\mu}$m at room temperature and decreases exponentially upon increasing the temperature at which the reconstruction takes place.

I.V.L. acknowledges the European Union MaX Center of Excellence (EU-H2020 Grant No. 824143). This work has been carried out using computing resources of the federal collective usage center Complex for Simulation and Data Processing for Mega-science Facilities at NRC “Kurchatov Institute” (http://ckp.nrcki.ru/).
\bibliography{reconstruction}
\end{document}